\documentclass[twocolumn,showpacs,preprintnumbers,amsmath,amssymb]{revtex4}

\usepackage{graphicx}
\usepackage{dcolumn}
\usepackage{bm}

\begin{document}
\def\al{&\!\!\!\!}
\def\x{{\bf x}}
\def\f{\frac}
\def\y{\frac{1}{2}}
\preprint{APS/123-QED}

\title{Dark companion of baryonic
matter}

\author{Y. Sobouti}
 \altaffiliation[ ]{Institute for Advanced Studies in Basic Sciences - Zanjan, Iran}
 \email{sobouti@iasbs.ac.ir}
\affiliation{%
Institute for Advanced Studies in Basic Sciences-Zanjan, Iran}
%

\date{\today}

\begin{abstract}
Flat or almost flat rotation curves of spiral galaxies can be
explained by logarithmic gravitational potentials. The field
equations of GR admit of spacetime metrics with such behaviors.
The scenario can be interpreted either as an alternative theory of
gravitation or, equivalently, as a dark matter paradigm. In the
latter interpretation, one is led to assign a dark companion to
the baryonic matter who's size and distribution is determined by
the mass of the baryons. The formalism also opens up a way to
support Milgrom's idea that the acceleration of a test object in a
gravitational field is not simply the newtonian gravitational
force $g_N$, but rather an involved function of $(g_N/a_0)$, $a_0$
MOND's universal acceleration.
\end{abstract}
\maketitle
\section{\label{sec:level1}Introduction}

The goal of the paper is to understand the idiosyncrasy of the
rotation curves of spiral galaxies. The newtonian or the GR
gravitation of the observable matter is not sufficient to explain
the large asymptotic speeds of test objects in orbits around the
galaxies, nor their slow decline with increasing distances. In
search of the missing gravity, alternative theories of gravitation
and/or of dark matter are proposed. In a recent work \cite{sob08}
we pointed out that no one has reported a case where there is no
baryonic matter, but there is a dynamical issue to be settled. We
argued that if the dark matter reveals itself only in the presence
of the baryonic one, it is logical to assume that the two are twin
companions. On the other hand, both dark matter scenarists and (at
least some) alternative theorists explain the rotation curves of
spirals equally satisfactorily. We argue, if two people give
correct answers to the same question, they ought to be saying the
same thing, albeit in different languages. And since in an
alternative theory one gives a definite rule for the gravity
field, there must be rules to govern the mutual companionship of
the dark and baryonic matters.

We begin with a GR formalism and show that spacetime metrics with
logarithmic behaviors are accommodated by Einstein's field
equations and can adequately explain the anomalous features of the
dynamics of the spirals. Conclusions are interpretable either in
terms of an alternative theory of gravitation, or as a dark matter
paradigm. With an advantage, however: the questions, how much dark
matter accompanies a given baryonic mass, how it is distributed,
and what is its equation of state, are also answered.

\section{ Model and  Formalism }

We are concerned with the outer reaches of spiral galaxies (a
baryonic vacuum), where the rotation curves display non classical
features. Their asymptotic speeds do not have a Keplerian
decline\cite{beg} and follow the Tully-Fisher relation\cite{tul}.
We approximate the galaxy by a spherically symmetric distribution
of baryonic matter.
The spacetime around it will accordingly
be spherically symmetric and static:
\begin{eqnarray}
ds^2=-B(r)dt^2+A(r)dr^2+r^2\left(d\theta^2+\sin^2\theta
d\varphi^2\right)
 \label{eq1}.
\end{eqnarray}
We adopt a dark matter language and assume that the galaxy
possesses a static dark perfect gas companion of density
$\rho_d(r)$, of pressure $p_d(r)<<\rho_d(r)$, and of covariant
4-velocities $U_t=-B^{1/2},U_i=0, ~i=r,\theta,\varphi$. Einstein's
field equations become.
\begin{eqnarray}
R_{\mu\nu}-\frac{1}{2}g_{\mu\nu}R=-T_{\mu\nu}=-[p_d
g_{\mu\nu}+(p_d+\rho_d)U_\mu U_\nu],
 \label{eq2}
\end{eqnarray}
where we have let $8\pi G=c^2=1.$ To respect the Bianchi
identities and the conservation laws of the baryonic matter, one
must have ${T^{\mu\nu}}_{\mu\nu}=0$. The latter, in turn, leads to
the hydrostatic equilibrium for the dark fluid, that is, if one
wishes to attribute such notions to a hypothetical entity.

From Eq. (\ref{eq2}) the two combinations
$R_{tt}/B+R_{rr}/A+2R_{\theta\theta}/r^2$ and $R_{tt}/B+R_{rr}/A$
give

\begin{eqnarray}
\al\al
\f{1}{r^2}\left[\f{d}{d r}\left(\f{r}{A}\right)-1\right]=-\rho_d,\label{eq3}\\
\al\al \f{1}{r A}\left(\f{B'}{B}+\f{A'}{A}\right)=\rho_d+p_d,
\label{eq4}
\end{eqnarray}
respectively. Neglecting $p_d$ in comparison with $\rho_d$ and
eliminating $\rho_d$ between the two equations gives
\begin{eqnarray}
\f{B'}{B}=\f{1}{r}(A-1). \label{eq5}
\end{eqnarray}
We now assume that $A(r)-1$ is a well behaved and differentiable
function of $r$ and has a series expansion in negative powers of
$r$,
\begin{eqnarray}
A-1=\sum_{n=0}\f{s_n}{r^n},~s_n~ \textrm{constant}. \label{eq6}
\end{eqnarray}
Substituting this expansion in Eq. (\ref{eq5}) and integrating the
resulting expression gives
\begin{eqnarray}
B=\left(\f{r}{r_0}\right)^{s_0}\exp\left({-\sum_{n=1}
\f{s_n}{nr^n}}\right)\approx\left[1+s_0\ln
\left(\f{r}{r_0}\right)-\f{s_1}{r}-\cdots\right]. \label{eq7}
\end{eqnarray}
We note that $s_0$ is dimensionless and $s_n,~n\geq1$, has the
dimension $\textrm{(length)}^n$. The right hand side expression is
the weak field approximation  and holds for  $s_n/r^n<<1,~
\forall~ n$.

With A and B known, the density $\rho_d$ can be calculated from
either of Eqs.~(\ref{eq3}) or (\ref{eq4}). Here, however, we adopt
a weak field point of view, $B(r)=1+2\phi_{grav}/c^2$, and
calculate $\rho_d$ from Poisson's equation. Thus,
\begin{eqnarray}
4\pi G\rho_d~~\al=\al~~\f{1}{2}
c^2\nabla^2(B-1)~=~\f{c^2}{2r^2}\left[s_0-\sum_{n=2}(n-1)\f{s_n}{r^n}\right].
\label{eq8}
\end{eqnarray}
Hereafter, we restore the physical dimensions $8\pi G$ and $c^2$
for clarity. The pressure of the companion fluid is obtained from
${T^{\mu\nu}}_{;\nu}=0$,
\begin{eqnarray}
\f{p_d'}{p_d+\rho_d}\approx \f{p_d'}{\rho_d}=-\f{1}{2r}(A-1).
 \label{eq9}
\end{eqnarray}
Integration is straight forward. The first two terms in the series
are
\begin{eqnarray}
p_d~~\al=\al~\f{c^2 s_0}{16\pi G r^2}\left[\f{1}{2}s_0+\f{ s_1}{3
r}\right].\label{eq10}
\end{eqnarray}
Upon elimination of $r$ between Eqs.~(\ref{eq8})~ and
~(\ref{eq10}) one obtains the equation of state, $p_d(\rho_d)$. It
is barotropic.
 We conclude this section by writing down the
dynamical acceleration of a test object circling the galaxy with
the speed $v$
\begin{eqnarray}
a_{\textrm{dyn}}~~\al=~\al\f{v^2}{r}=\f{1}{2}c^2 B'
=\f{1}{2}c^2\left[\f{s_0}{r}+\f{s_1}{r^2}+\cdots+\f{s_n}{r^{n+1}}+\cdots\right].
\label{eq11}
\end{eqnarray}
\section{What are ${s_n}'\rm s$ }
The $s_1$ term in Eqs.~(\ref{eq6})-(\ref{eq11}) represents the
classic gravitation of the baryonic matter with a force range of
$r^{-2}$. Magnitude-wise, $s_1$, should be identified with the
Schwarzschild radius of the spherical galaxy, $s_1=2GM/c^2$. The
$s_0$-term is not a classical term. It has a force range $r^{-1}$
and dominates all other terms at large distances. It is
responsible for the large asymptotic speeds and their non
Keplerian decline at far reaches of the spirals.  In \cite{sob07}
and \cite{sob08} we resorted to the Tully-Fisher relation (the
proportionally of the asymptotic speed ,
$v_\infty=c(s_0/2)^{1/2}$, to the fourth root of the mass of the
host galaxy) and arrived at
\begin{eqnarray}
s_0=\alpha\left(\f{M}{M_\odot}\right)^{1/2},~\alpha~\textrm{constant}.
\label{eq12}
\end{eqnarray}
In  weak accelerations  (less than certain `universal
acceleration' $a_0$), Milgrom's MOND \cite{mil} anticipates a
force field  $(a_0 g_N)^{1/2}$, instead of the newtonian
gravitation, $g_N=GM/r^2$.
 The far distance limit
of Eq.~(\ref{eq11}) with $\alpha$ given by Eq.~(\ref{eq12}) is  of
Milgrom's form. Comparing the two formalisms, one finds
$\alpha=2(a_0 GM_\odot)^{1/2}c^{-2}$. Either from this expression,
with $a_0\approx 1.2\times 10^{-8}\textrm{cm/sec}^2$ \cite{beg},
or from a direct statistical analysis of the asymptotic speeds of
spirals \cite {sob07} one finds
\begin{eqnarray}
\alpha \approx 2.8\times 10^{-12},\textrm{ dimensionless
`universal constant'}. \label{eq13}
\end{eqnarray}
The remaining $s_n$-terms, $n\geq2$, in
Eqs.~(\ref{eq6})-(\ref{eq11}) are also nonclassical. The range of
their force is $r^{-(n+1)}$ (not to be confused with the multipole
fields of extended objects). There is no compelling observational
evidence for their existence in regions external to  a spherical
distribution of matter. Nevertheless, we retain them for a
possible formal support they may give to Milgrom's MOND, to be
elaborated below.

\textbf{A conjecture}: There is a surprise in Eq.~(\ref{eq11}).
Upon elimination of $r$ in favor of $g_N=GM/r^2$, one may write it
as
\begin{eqnarray}\label{eq14}
\al\al\f{a_{\textrm{dyn}}}{a_0}=\left(\f{g_N}{a_0}\right)^{1/2}+
\left(\f{g_N}{a_0}\right)
+\cdots+\alpha_n\left(\f{g_N}{a_0}\right)^{(n+1)/2}+\cdots.
\end{eqnarray}
where $\alpha_n$'s can be expressed in terms of $s_n$'s through a
term-by-term comparison of Eqs. (\ref{eq11}) and (\ref{eq14}). One
obtains
\begin{eqnarray}
\al\al \alpha_n=\f{c^2
s_n}{2a_0}\left(\f{a_0}{GM}\right)^{(n+1)/2}, n=2,3,\cdots,
 \label{eq15}\end{eqnarray}
 or
 \begin{eqnarray}
 \al\al
 s_n=\f{2a_0}{c^2}\alpha_n\left(\f{GM}{a_0}\right)^{-(n+1)/2}.\label{eq16}
\end{eqnarray}
All $\alpha_n$'s are dimensionless. Apparently, Eq. (\ref{eq14})
is an expansion of the dynamical acceleration in a power series of
$(g_N/a_0)^{1/2}$. The coefficient of the first term is the
`universal constant' 1 because of the  `universal' Tully-Fisher
relation. The coefficient  of the second term is 1 because of the
universal law of newtonian gravitation in the weak field regime.
Now the conjecture: If there is any significance attached to the
series expansion of Eq.~(\ref{eq14}) beyond the first two terms,
is it
possible that in the remaining terms \\

``All $\alpha_n$'s are universal constants (not necessarily 1),
and independent from the mass of the host baryonic matter centered
at the origin"?\\

The proof or disproof of the conjecture should come from
observations. We recall, however, Milgrom's  stand  that the
dynamical acceleration of a test body is not simply proportional
to $g_N$, but it is a involved function of $g_N/a_0$, and vice
versa, $g_N$ is a function of $a_{dyn}/a_0$. His suggestion for
this function is through an, almost arbitrary, interpolating
function.  If the conjecture above holds, Eq.~(\ref{eq14}) can be
considered as a series expansion of one such function, and a
support for Milgrom's idea.

\section{Concluding remarks}

That logarithmic potentials are natural solutions of Einstein's
field equations is the highlight of the paper. They enable one to
arrive at a law of gravitation alternative to that of Newton
and/or to those known to GR. Equivalently, one may choose to
attribute dark companions to baryonic matters. In the case of a
spherically symmetric baryonic mass, the size and distribution of
the density and pressure of the companion, outside the baryonic
mass, are given by Eqs. (\ref{eq8}) and (\ref{eq10}).

The spacetime is a baryonic vacuum but not a dark matter one. The
consequences are noteworthy. For example:\\ The spacetime is not
flat. Contraction of Eq.~(\ref{eq2}) gives
$$R=-(3p_d+\rho_d)\approx-\f{s_0}{r^2}+O\left(r^{-4}\right).$$
 The 3-space is not flat. Direct calculation with $
g^{(3)}_{ij},i,j=r,\theta,\varphi$, yields
$$R^{(3)}=-\f{2}{r^2}\f{d}{d r}(r \rho_d)\approx-2\f{s_0}{ r^2}+
O\left(r^{-4}\right).$$
 There is an excess lensing \cite{men07}. Contribution from the $s_0$ term alone is
$$\delta\beta=\f{1}{2}\pi s_0.$$
Due to the smallness of both $s_0$ and Sun's mass, effects in the
scale of the solar system are immeasurably small\cite{sob08}.

That the dynamical acceleration of a test object in the external
gravitational field of a spherical mass could have a series
expansion in $(g_N/a_0)$, in accord with Milgrom's idea, is an
intriguing idea. The support for it should come from observations.

A word of caution: The paper relies heavily on observations
pertaining to spiral galaxies. Its conclusions may be scale
dependent, not applicable to systems with scales larger than
galactic scales. In a recent paper Bernal et al \cite{men} analyze
weak lensing data from clusters of galaxies on the basis of the
metric field of \cite{sob07} (similar to those of Eqs.~\ref{eq6}~
and \ref{eq9}~). They conclude, in the notation of this paper,
$s_0\propto M^{1/4}$, instead of $M^{1/2}$ of Eq.~(\ref{eq12}).
This finding while raises an alarm against extrapolation to larger
systems, clusters of galaxies and beyond, at the same time opens
the question that deviations from the newtonian or GR gravitations
may have a hierarchical structure depending on the size of the
system under study.

Shortcomings of the paper and the open questions it leaves behind
should also be mentioned. The theory developed here is for a
spherical distribution of mass. Extension to extended objects and
to many body systems is not a trivial task. It may require further
assumptions not contemplated so far. The difficulty lies in the
facts that a) the added $s_0$- and $s_n$- terms, $n\geq2$, are not
linear in the mass of the baryonic matter. The nonlinearity is
much more complicated than that of GR. b) In the parlance of a
dark matter paradigm, the dark companion of a localized baryonic
matter is not localized and extends to infinity. As a way out of
the dilemma, we are planning to expand an extended object into its
localized monopole and higher multipole moments, and see if it is
possible to find a dark multipole moment for each baryonic one,
more or less in the way done for the monopole moment.\\

\textbf{Acknowledgment}:  The author wishes to acknowledge a
discussion with Sergio Mendoza that eventually lead to the
expansion of Eq. (\ref{eq14}).

\end{document}